\documentstyle[11pt,a4]{article}
\begin{document}



\def\ga{\alpha}
\def\gb{\beta}
\def\gc{\gamma}
\def\gcp{\gamma^\prime}
\def\gd{\delta}
\def\gep{\epsilon}
\def\gl{\lambda}
\def\gL{\Lambda}
\def\gk{\kappa}
\def\go{\omega}
\def\gp{\phi}
\def\gs{\sigma}
\def\gt{\theta}
\def\gC{\Gamma}
\def\gD{\Delta}
\def\gO{\Omega}
\def\gT{\Theta}


\def\be{\begin{equation}}
\def\ee{\end{equation}}
\def\ba{\begin{array}}
\def\ea{\end{array}}
\def\bea{\begin{eqnarray}}
\def\eea{\end{eqnarray}}
\def\bes{\begin{eqnarray*}}
\def\ees{\end{eqnarray*}}
\def\bsea{\begin{subeqnarray}}
\def\esea{\end{subeqnarray}}
\def\btab{\begin{tabular}}
\def\etab{\end{tabular}}
\def\lra{\longrightarrow}
\def\lms{\longmapsto}
\def\ra{\rightarrow}
\def\ms{\mapsto}
\def\pa{\partial}
\def\ll{\parallel}


\newcounter{subequation}[equation]

\makeatletter
\expandafter\let\expandafter\reset@font\csname reset@font\endcsname
\newenvironment{subeqnarray}
  {\def\@eqnnum\stepcounter##1{\stepcounter{subequation}{\reset@font\rm
      (\theequation\alph{subequation})}}\eqnarray}%
  {\endeqnarray\stepcounter{equation}}
\makeatother

\newtheorem{lemma}{Lemma}
\newtheorem{satz}{Satz}
\newtheorem{theorem}{Theorem}
\newtheorem{cor}{Corollary}
\newtheorem{define}{Definition}


\newcommand{\N}{\mbox{I\hspace{-.4ex}N}}
\newcommand{\C}{\mbox{$\,${\sf I}\hspace{-1.2ex}{\bf C}}}
\newcommand{\Cs}{\mbox{$\,${\sf I}\hspace{-1.2ex}C}}
\newcommand{\Z}{\mbox{{\sf Z}\hspace{-1ex}{\sf Z}}}
\newcommand{\R}{\mbox{\rm I\hspace{-.4ex}R}}
\newcommand{\1}{\mbox{1\hspace{-.6ex}1}}


\def\la{\wi{\mbox{Tr} J^2}}
\def\cp{c^\prime}
\def\rum{r_{\mu_{\rm min}}}
\newcommand{\me}{\mbox{e}}
\newcommand{\wum}{\left(1 + \frac{\tau^2}{\omega^2}\right)^{-\frac{1}{2}}}
\newcommand{\wup}{\left(1 + \frac{\tau^2}{\omega^2}\right)^{\frac{1}{2}}}
\newcommand{\ta}{\tilde{a}}
\newcommand{\tJ}{\tilde{J}}
\newcommand{\DI}[1]{\mbox{$\displaystyle{#1}$}}
\newcommand{\HH}{\mbox{I\hspace{-.4ex}H }}
\newcommand{\wi}[1]{\widehat{#1}}
\newcommand{\SL}{\mbox{SL(2,\rm I\hspace{-.4ex}R)}}
\newcommand{\SLs}{\mbox{SL(2,\rm I\hspace{-.4ex}R) }}
\newcommand{\slalgs}{\mbox{sl(2,\rm I\hspace{-.4ex}R) }}
\newcommand{\slalg}{\mbox{sl(2,\rm I\hspace{-.4ex}R)}}
\newcommand{\scoset}{\mbox{SL(2,\rm I\hspace{-.4ex}R)/SO(2)}}
\newcommand{\scosets}{\mbox{SL(2,\rm I\hspace{-.4ex}R)/SO(2) }}



\thispagestyle{empty}

\hbox to \hsize{%
  \vtop{\hbox{accepted for publication in }\hbox{\sl Phys. Lett. B}} \hfill
  \vtop{\hbox{MPI-PhT/96-69}}}

\vspace*{1cm}

\bigskip\bigskip\begin{center}
{\bf \Huge{Group Theoretical Quantization of Schwarzschild 
and Taub-NUT}}
\end{center}  \vskip 1.0truecm
\centerline{{\bf Helia Hollmann\footnote{e-mail: 
hollmann@mppmu.mpg.de}}}
\vskip5mm
\centerline{Max-Planck-Institut f\"{u}r
 Physik, Werner-Heisenberg-Institut}
\centerline{F\"ohringer Ring 6, 80805 Munich, Germany}
\vskip 2cm
\bigskip \nopagebreak \begin{abstract}
\noindent
Stationary spherically symmetric gravity is equivalent to a 
nonlinear coset sigma model on \scosets coupled to a 
gravitational remnant. Classically there are stationary 
solutions besides the static Schwarzschild metric labeled 
by the Schwarzschild mass $m$ and the Taub-NUT charge $l$. 
Imposing the \SLs symmetry at the quantum level the Wheeler-DeWitt 
equation becomes related to the Casimir operator on the coset, which 
makes the system amenable to exact quantization.

\end{abstract}

\newpage\setcounter{page}1


The quantization of Einstein's gravity is one of the outstanding
and most challenging areas of theoretical physics. 
Freezing all but a few of the infinitely many degrees of
freedom of the gravitational field, solutions for special sectors of
Einstein's vacuum theory are within reach \cite{KasThi93}, 
\cite{Kuc94}, \cite{MarGegKun94}, \cite{CavdeAFil951}.
It is believed that at least some relevant features of the 
quantum theory are preserved.
Even  restricting oneself to a particularly symmetric case, that is
to stationary spherical symmetry --- as it is done here --- 
the space of classical solutions contains black holes. They are 
in some respect the true particles of general relativity: 
coupling a point particle 
to its own gravitational field turns it into a black hole 
\cite{Dam86}. Furthermore classical black holes are stable under 
local perturbations \cite{Cha83}, and they have remarkable 
uniqueness properties \cite{BreMaiGib88}. This is why they 
are considered to be the fundamental pieces of a quantum theory 
of gravitation, and the investigation of the spherically symmetric 
sector of Einstein's theory of gravitation is seen in a different 
light.  

This paper mainly splits into a classical and a quantum part. In 
the beginning the spherically symmetric sector of Einstein's
gravity is introduced and a Kaluza-Klein like reduction is
performed \cite{BreMai86}. Classically the theory can be solved,  
and the physical parameters turn out to be the Taub-NUT charge $l$ 
and the Schwarzschild mass $m$. In the following the problem is 
treated as a constrained system. The observables are identified and 
shown to form the Lie algebra \slalg. The Hilbert space of the 
model is built from the unitary irreducible representations of 
the \SLs symmetry group. The quantum mechanical model can be solved 
completely. 

One starts with the following parametrization of the four 
dimensional line element
\[
  ds^2 = \tau (dt - B_m dx^m)^2 - \tau^{-1} h_{mn} dx^m dx^n.
\] 
As in Kaluza-Klein theories the existence of a timelike 
Killing vector field can be used to perform a dimensional reduction 
\cite{Kal21} of Einstein's gravity. It leads to 
three dimensional gravity coupled to an abelian vector field 
and a scalar. After elimination of the field strength $F_{mn}$ 
of $B_m$ by the Bianchi identity 
\[
  \tau^2 F_{mn} = \gep_{mnp} \pa^p \go
\]
one arrives at a nonlinear \scosets sigma model 
coupled to gravity:  
\be
\label{lag}
  {\cal L} = \sqrt{h}
  \left(
    - \frac{R}{2} ~+~ \frac{h^{mn}}{8} \mbox{\rm Tr}
    \left(
      \chi^{-1} \pa_m \chi~\chi^{-1} \pa_n \chi
    \right)
  \right),
\ee
where
\[
  \chi =
  \left(
    \ba{cc}
      \tau + \DI{\frac{\go^2}{\tau}} &  \DI{\frac{\go}{\tau}} \\
      & \\
      \DI{\frac{\go}{\tau}}  & \DI{\frac{1}{\tau}}  \\
    \ea
  \right).
\]
$h_{mn}$ denotes the three dimensional metric, $\sqrt{h}$ the square 
root of the determinant, $R$ is the scalar curvature associated 
with $h_{mn}$, and $\chi$ is an element of the Riemannian space 
\scoset. 
$\tau$ is the norm of the Killing vector and plays 
the role of a gravitational potential. $\go$ is the so called 
twist or NUT potential, which can also be considered as a kind of 
``magnetic potential'' in analogy with Maxwell's theory. 
To further simplify the problem one imposes an additional symmetry 
on the remaining three spatial dimensions: 
\[
  h_{mn} dx^m dx^n \:=\: N^2 d\rho^2 + f^2 d\gO^2.
\] 
Then the Lagrangian (\ref{lag}) simplifies to 
\[
  {\cal L} = N \: 
    \left( 
      \frac{f^{\prime^2}}{N^2} + 1 - \frac{f^2}{4 N^2 \tau^2}
      ( \tau^{\prime^2} + \go^{\prime^2} )
    \right).
\]
The additional SO(3) symmetry  reduces the equations of motion 
to ordinary nonlinear differential equations \cite{Tau51}, 
\cite{DobMai82}, which can be integrated. The space of 
classical solutions consists of the Schwarzschild metric and a 
one parameter family connecting the Taub-NUT solution with the 
Schwarzschild solution. The geometrical difference between 
these is that the latter admits an isometry group whose action 
on four dimensional space-time has three dimensional orbits, 
whereas they are 2-spheres for the Schwarzschild solution. 
Another interpretation is given in terms of the 
difference between staticity and stationarity \cite{BreMaiGib88}: 
In the static case there are two symmetries: a time translation symmetry 
and a time reflection symmetry. For Taub-NUT the fields being time 
translation invariant fail to be time reflection invariant: 
the adjacent orbits of the Killing vector fields twist around 
each other! Both solutions are asymptotically flat, but only the 
Schwarzschild solution is asymptotically Minkowski. 
The space of classical solutions, which consists of the 
Schwarzschild solution only, is the static truncation GL(1) of the 
original sigma model \scoset. 

A suitable choice of coordinates leads to 
\[
  \go = \frac{l}{\rho} + O(\frac{1}{\rho^2})
  \quad \mbox{and} \quad
  \tau = 1 - \frac{2m}{\rho} + O(\frac{1}{\rho^2}). 
\]
The coefficients $l$ and $m$ of the asymptotic expansion of $\go$ and 
$\tau$ are called the Taub-NUT charge $l$ and the Schwarzschild mass 
$m$. They are the ``physical'' parameters of the space of classical 
solutions. Within the sigma model formalism the currents which 
are in this case also ``global charges'', are contained in the matrix 
\[
  J = \frac{f^2}{2 N^2} \: \chi^{-1} \chi^\prime  
    = \chi_0^{-1} \chi_1, 
\]
where $\chi_0$ and $\chi_1$ are the coefficients of order zero and 
one of the asymptotic multipole expansion of the \scosets matrix $\chi$. 
After a suitable reparametrization one of the equations of motion 
becomes the equation of geodesic motion in the coset space 
\scoset. This yields  
a geometrical interpretation of the Taub-NUT charge $l$ and 
the Schwarzschild mass $m$. The requirement that $\chi$ tends to the 
``vacuum'' configuration in the limit $\rho \ra \infty$ fixes the 
values of $\go$ and $\tau$ at the starting point of the geodesic. 
Apart from this there exists a uniquely determined velocity vector 
$(\dot{\tau}_\infty, \dot{\go}_\infty) = (m, l)$  which 
characterizes the geodesic completely. 

The gravitational remnant parametrized by the coordinate $f$ decouples 
from the sigma model and can be solved separately. 

The quantization of the spherically symmetric sector of Einstein's 
gravity is achieved via the canonical ADM quantization. 
At the classical level the ADM formalism selects the coordinates 
for which the Hamiltonian itself becomes a constraint. The four 
dimensional space-time is split into a three dimensional spacelike 
hypersurface and a time. A lapse function $N$ and a shift function 
$N_i$ are introduced to describe how the 3D hypersurface 
at time $x_0=$ constant with a given 3-metric and an extrinsic 
curvature is related to the hypersurface at time $x_0 + \gd x_0$. 
The lapse and shift functions enter the Hamiltonian formalism 
as Lagrange multipliers. In this paper a modified Hamiltonian formalism 
is to be implemented. The slicing is performed according to  
$\rho$, which is a spacelike coordinate. The formalism can be found 
in \cite{Mar94}. Due to spherical symmetry the lapse function is the 
only Lagrange multiplier which survives. In other words: it expresses 
the invariance under $\rho$-reparametrization and leads to the only 
primary first class constraint of the theory, the Hamiltonian 
\[
  H = \frac{1}{4} \pi_f^2 - 1 
     - \frac{\tau^2}{f^2} (\pi_\tau^2  + \pi_\go^2),
\]
where $\pi_\go, \pi_\tau$ and $\pi_f$ denote the canonical 
$\rho$-momenta. $H$  generates the gauge transformations. 
Observables are functions on the constraint surface which are gauge 
invariant, i.e. their Poisson brackets with the first class 
constraints vanish. By definition they do not evolve in ``time'' and 
therefore there is a one to one correspondence between the 
observables and the initial data. Expansion of the sigma model current 
\bea
  J &=&  
  \left(
    \ba{ll}
      - J^0 & J^+ \\
      J^- & J^0
    \ea
  \right) \\
  &=&
  \left(
    \ba{cc}
      - \tau \pi_\tau - \go \pi_\go & - \pi_\go \\
      \tau^2 \pi_\go - 2 \go \tau \pi_\tau - \go^2 \pi_\go &
      \tau \pi_\tau + \go \pi_\go
    \ea
  \right) \nonumber
\eea
into a suitable Lie algebra basis yields two observables $J^0$ and 
$J^+$ of the theory, which commute with $H$:   
\[
  \{ H, J^0 \} = 0, \qquad \{ H, J^+ \} = 0.
\]
Evaluating $J^0$ and $J^+$ on the classical solutions gives  
the Schwarzschild mass $m$ and the Taub-NUT charge $l$ respectively. 
The third coefficient $J^-$ is not linearly independent.
The currents $J^0$, $J^+$ and $J^-$ form an \slalgs algebra which becomes 
important in the quantization process. The finite transformation related 
to $J^0$ is a scale transformation $\go \ms \me^\ga \go$, 
$\tau \ms \me^\ga \tau$. To $J^+$ corresponds a shift in $\go$.

In the sequel a Dirac quantization of this sector of Einstein's 
gravity is performed. In the Schr\"{o}dinger representation the 
fields are turned into multiplication operators $\Phi \ms \wi{\Phi}$ 
and the momenta become differentiation operators 
$\pi_\Phi \ms -i \pa_\Phi$, where $\pa_\Phi$ is an abbreviation 
for $\frac{\pa}{\pa_\Phi}$:
\bes
  \wi{J^0} &=& -i \: \tau \pa_\tau \:-\: i \: \go \pa_\go \\
  \wi{J^+} &=& i \: \pa_\go \\
  \wi{J^-} &=& i \: (\tau^2 \:-\: \go^2) \: \pa_\go
     \:-\: 2 i \: \go \tau \: \pa_\tau \\
  \la &=& - 2 \: \tau^2 \:(\pa_\tau^2 \:+\: \pa_\go^2) \\
  \wi{H} &=& - \frac{\pa_f^2}{4} \:+\: \frac{1}{2 f^2} \:
     \la \:-\: 1.
\ees
Applied to a wave function $\psi$ the Hamiltonian defines the so 
called Wheeler-DeWitt equation $\wi{H} \psi = 0$.
The following commutation relations hold:
\[
  [\: \wi{H}, \: \wi{J^0}] \:=\: 0 \qquad
  [\: \wi{H}, \: \wi{J^+}] \:=\: 0 \qquad
  [\: \wi{H}, \: \wi{J^-}] \:=\: 0,  
\]
i.e. the current operators are observables. The Casimir operator 
$\la$ commutes with the Hamiltonian and the currents:
\[
  [\: \la, \: \wi{H}] \:=\: 0 \qquad
  [\: \la, \: \wi{J^0}] \:=\: 0 \qquad
  [\: \la, \: \wi{J^+}] \:=\: 0 \qquad
  [\: \la, \: \wi{J^-}] \:=\: 0,  
\]
and $\wi{J^0}, \wi{J^+}, \wi{J^-}$ form an \slalgs algebra:
\[
  [\: \wi{J^+}, \: \wi{J^0}] \:=\: - i \wi{J^+}, \qquad
  [\: \wi{J^+}, \: \wi{J^-}] \:=\: 2 i \: \wi{J^0}, \qquad
  [\: \wi{J^0}, \: \wi{J^-}] \:=\: - i \wi{J^-}.
\]
The Casimir operator $\la$ is an essential part of the Wheeler-DeWitt 
equation. As it commutes with the current operators, $\la$ and any one 
of the $J$'s can be ``measured simultaneously'', i.e. they can be 
simultaneously diagonalized. $\wi{J^0}, \wi{J^+}$ and $\wi{J^-}$ do not 
commute with each other. Therefore even formally a direct 
``measurement'' of the Schwarzschild mass and the Taub-NUT charge 
is not possible. 
The interpretation of the current operators as observables 
of the theory strongly suggests to preserve the \SLs symmetry 
at the quantum level. That is, the quantum mechanical Hilbert 
space is built from the unitary irreducible representations of the group 
\SL. Finding a solution of the Wheeler-DeWitt equation becomes therefore 
basically a group theoretical problem as the equation  splits into 
an $f$ dependent and an $f$ independent part, which is the Casimir 
operator on the group up to a constant term.
The solutions of the eigenvalue equation for 
$\la$ yield a complete set of eigendistributions in the sense that 
they are a basis of the representation space.   

There are two particularly useful possibilities to diagonalize 
the Casimir operator $\la$. 
On one hand one solves the differential equations for $\la$ and the Taub-NUT 
charge operator $\wi{J^+}$ simultaneously:  
\bsea
  \label{dgl1}
  - \tau^2 \left( \pa_\tau^2  \:+\: \pa_\go^2 \right) \:
    \psi_{\gl L}(\go, \tau)
  & = & \gl \: \psi_{\gl L}(\go, \tau), \\
  i \: \pa_\go \: \psi_{\gl L}(\go, \tau)
  & = & L \: \psi_{\gl L}(\go, \tau).
\esea
This system of differential equations yields 
\[
  \psi_{\gl L}(\go, \tau) \:=\: \me^{-i L \go} \: \sqrt{\tau} \: 
  \left[ 
    C_1 \: \mbox{I}_k(|L| \tau) \:+\: C_2 \: \mbox{K}_k(|L| \tau)
  \right], 
\]
where $k = \sqrt{1 - 4 \gl}$. I$_k$ and $K_k$ denote the Bessel 
functions of first and second kinds with imaginary index $k$.
On the other hand one can diagonalize the Schwarzschild mass 
operator $\wi{J^0}$ and the Casimir operator $\la$
\bsea
  \label{dgl2}
  - \tau^2 \: ( \pa_\tau^2 \:+\: \pa_\go^2 ) \:
  \psi_{\gl M}(\go, \tau)
  &=& \gl \: \psi_{\gl M}(\go, \tau) \\
  - i \: \left( \tau \pa_\tau \:+\: \go \pa_\go \right) \:
  \psi_{\gl M}(\go, \tau)
  &=& M \: \psi_{\gl M}(\go, \tau), 
\esea
which yields 
\be
\label{le1}
  \psi_{\gl M}(\go, v) \:=\: \psi_{\gl M}^1(\go, v)
      \:+\: \psi_{\gl M}^2(\go, v)
\ee
with
\bsea
  \psi_{\gl M}^1(\go, v) \!\!\!\!\! &=&\!\!\!\!\!
     C_1 \: \go^{i M}  \sqrt{v} 
     \left( 1 + v^2 \right)^{\frac{i M}{2} - \frac{1}{4}}
     \mbox{P}^{-\frac{k}{2}}_{-i M -\frac{1}{2}}
     \left( \frac{1}{\sqrt{1+v^2}} \right), \nonumber \\
  \psi_{\gl M}^2(\go, v) \!\!\!\!\! &=&\!\!\!\!\!
     C_2 \: \go^{i M} v^{k + \frac{1}{2}} |v|^{-k} 
     \left( 1 + v^2\right)^{\frac{i M }{2} - \frac{1}{4}}
     \mbox{P}^{\frac{k}{2}}_{-i M  - \frac{1}{2}}
     \left( \frac{1}{\sqrt{1 + v^2}} \right)
     \nonumber
\esea
and $v = \go/\tau$ and $\gl \neq  \frac{1}{4} - n^2,~ n \in \Z$.
P$^{\pm \frac{k}{2}}_{-iM-\frac{1}{2}}$ are the associated 
Legendre functions of first kind.  

There is a one to one correspondence between the coset 
\scosets and the Poincar\'e upper half plane 
$\HH := \{\go + i \: \tau \in \C ,~ \tau > 0\}$. 
Taking a function of the dense subspace  
$C_c^\infty(\HH)$ of the representation space $L^2(\HH, d\mu)$, 
where the group invariant measure is $d\mu = d\go d\tau/\tau^2$, the 
current operators $\wi{J^+}$ and $\wi{J^0}$ are shown to be symmetric. 
The current operators $\wi{J^+}$ and $\wi{J^0}$ turn out to be 
already essentially 
self-adjoint. No constraints on the spectra of $\wi{J^+}$ and 
$\wi{J^0}$ arise: it consists of the whole real line for each operator. 
Negative 
Taub-NUT charges as well as negative masses belong to the spectrum, 
too. Suitable solutions are those which belong to the continuous 
spectrum of the operators. 
The Bessel functions of first kind with imaginary argument  
grow exponentially at infinity and do not contribute to the spectral 
decomposition of the Laplacian.  
Concerning the second system of differential equations,  
$\psi^2_{\gl M}$ drops out because of discontinuities on the imaginary 
axis. 

The ``time'' dependent part of the Wheeler-DeWitt equation is an ordinary 
differential equation in $f$:
\be
\label{WdWt}
  \psi^{\prime \prime} \:+\:
  \left( 4 \:-\: \frac{2 \gl}{f^2} \right) \: \psi \:=\: 0.
\ee
The prime $^\prime$ denotes differentiation with respect to $f$.  
The dynamical state of the system is entirely hidden in the entities 
described by $\go$ and $\tau$. In contrast to the coordinate $\rho$ 
which was arbitrarily chosen, the coordinate $f$ has a geometrical 
meaning, as $1/f$ is the curvature of the 2-spheres, and serves as 
a geometrically defined ``time''. Therefore (\ref{WdWt}) describes 
the ``time'' evolution of the system. Because it is 
a second order differential equation, one cannot construct a 
positive semidefinite probability density that is conserved in 
``time''. But as in the case of the Klein-Gordon equation 
\cite{FesVil58} a conserved density
\[
  \rho  \:=\: \psi^\star \pa_f \psi \:-\: \psi \pa_f \psi^\star 
\]
can be found. 
One can interpret the $f$ dependent term in the differential 
equation (\ref{WdWt}) as a ``time dependent'' perturbation. It 
is instructive to transform the unperturbed differential equation 
into a system of first order equations $\pa_f \Psi = H_0 \Psi, \: 
\Psi = (\phi_0, \chi_0)^T$. For $2 \,\phi_0 = \pa_f \psi_0 + 2 i \, \psi_0$ 
and $2 \, \chi_0 = \pa_f \psi_0 - 2 i \, \psi_0$ the operator $H_0$ is 
diagonal, and the density $\rho_0$ reads: 
$\rho_0 = i ( \phi_0^\star \phi_0 - \chi_0^\star \chi_0)$. It is a 
difference of two positive semidefinite parts which correspond to 
two equivalent states.    
Given $\gl$ the solution of (\ref{WdWt}) approaches the free solution 
in the limit $f \ra \infty$ leading to a suggestion how to define 
two independent but equivalent states in this case, too. 
The differential equation is of Bessel type, and the 
appropriate linear combinations turn out to be the Bessel functions 
of third kind H$_\nu^{(1)}$ and H$_\nu^{(2)}$ 
\[ 
  \psi(f) \:=\: C_1 \: \sqrt{f} \: \mbox{H}_\nu^{(1)}(2 f) 
    \:+\:  C_2 \: \sqrt{f} \: \mbox{H}_\nu^{(2)}(2 f), \qquad 
    \nu = \frac{1}{2} \sqrt{ 1 + 8 \gl}.
\]
Using the Wronskian it can be shown that the density is again 
a difference of two positive semidefinite parts. 

This paper concludes with some remarks. There is a nice relation  
between the quantization of four dimensional reduced spherically 
symmetric gravity in its dual representation and a \SLs WZNW model 
in the point particle version, which is further reduced to a 
Liouville theory \cite{Ful96}. One ends up with a Hamiltonian 
system with constraints, which correspond to initial values. 
Choosing these initial values, the Hamiltonian of the latter system 
has exactly the form of the remaining part of the Wheeler-DeWitt 
equation (\ref{WdWt}), where the initial values take the place of 
the eigenvalue of the Casimir operator. The Liouville Hamiltonian is 
interpreted to be the Hamiltonian of a relativistic particle which 
moves in a potential. Formally, there are three cases to be 
distinguished, depending on the sign of the constant $\gl$ for the 
initial values. If $\gl>0$, the particle is affected by an infinitely 
high potential barrier, when it travels in the negative $f$-direction. For 
$\gl = 0$ the particle is free, and for $\gl < 0$ an infinitely deep 
potential valley attracts the particle towards the negative $f$-direction. 
Here $\gl>0$ holds, because the Casimir operator can be shown to be the 
square of the Taub-NUT charge and the Schwarzschild mass operator. 

There are also some important differences between these models.  
In F\"{u}l\"{o}p's paper \cite{Ful96} the third coordinate $f$ has 
an intrinsic group theoretical  
meaning leading to an interpretation of the potential. $\gl > 0$ 
corresponds to the continuous and $\gl<0$ to the discrete series of 
representation theory. F\"{u}l\"{o}p considers the momenta of the 
particle to be fundamental. As the physical meaning is contained 
in the currents, there is no reason for the momenta to  
become hermitian operators. Moreover it turns out that starting  
with hermitian momenta the currents are non-hermitian. On the other 
hand, the conditions that $\wi{J^0}, \wi{J^+}$ and $\wi{J^-}$ are 
hermitian defines these operators to be Lie derivatives.  
This also fixes the operator ordering of the Laplacian and 
therefore of the Hamiltonian, too. 

The more technical aspects of this paper are treated in a separate 
forthcoming publication.       
  
\section*{Acknowledgment}\addcontentsline{toc}
{section}{Acknowledgment}

The author is indebted to P. Breitenlohner and D. Maison for
numerous discussions on the subject. 


\end{document}